\documentclass[prl,twocolumn,aps,amssymb]{revtex4}
\usepackage{graphicx}
\usepackage{bbm}
\begin{document}

\noindent
{\bf Comment on ``Shadowability of Statistical Averages in Chaotic Systems''}\\
\newline
\indent
Lai {\it et al.} \cite{Lai:2002}  investigate systems 
with a stable periodic orbit and a coexisting chaotic saddle.
They claim that, by adding white Gaussian noise (GN), averages 
change with an algebraic scaling law above a certain noise
threshold and argue that this leads to a breakdown of shadowability 
of averages. Here, we show that (i) shadowability is not well defined 
and even if it were, no breakdown occurs. We clarify 
misconceptions on (ii) thresholds for GN. We further point out
(iii) misconceptions on the effect of noise on averages. Finally,
we show that (iv) the lack of a proper threshold can lead to any 
(meaningless) scaling exponent.  
\\
\indent
(i) Shadowing deals with macroscopic error propagation of noise 
bounded by a small value (e.g. computer accuracy of $ 10^ {-16}$), 
comparing a noisy and a `true' trajectory \cite{Grebogi:1990}.
Since the authors use GN, one cannot properly speak of shadowing. 
However, even for bounded noise, their claims are still doubtful.
In a recent work, the error of an {\it average} was found to be 
amplified by a factor of up to $10^{12}$, although the trajectory 
never exits the attractor \cite{Sauer:2002}. This renders a reliable 
computation truly unfeasible. In \cite{Lai:2002}, though, the effect 
on the averages is only of the same order as the variation of the 
noise level $D$.  
Furthermore, the averages in Figs. 1a and 2a in \cite{Lai:2002}
do change even less below $D=10^{-2.5}$. Therefore, shadowability is not
compromised at all.     
\\
\indent
(ii) The authors of \cite{Lai:2002} mention a threshold for GN 
above which the periodic orbit and the chaotic saddle would become connected. 
Yet, such a threshold does not exist, since the mean first exit time 
$\langle \tau \rangle$ is given by Kramers' law $\langle \tau \rangle \sim
\exp(\frac{\Delta U}{D})$, where $D$ is the noise variance and $\Delta U$ is 
either the potential \cite{Kramers:1940} or, for nonequilibrium and chaotic
systems, the quasipotential difference \cite{Graham:1984}. Thus, $\langle 
\tau \rangle$ varies with $D$, yielding a {\it different average for
every noise level}.  
\\
\indent
(iii) Generally, averages depend on the noise {\it for all noise levels}, 
implying that {\it no threshold can exist}, even with only {\it one} 
metastable state. 
For the linear map $x_{n+1} = a x_n + b + \xi_n$ with a fixed point 
$x_{\star}= \frac{b}{1-a}$ and white GN 
one gets  $\langle x ^2 \rangle = x_{\star}^2 + \frac{1}{1-a^2} \, D$. 
This is pictured in Fig. 1a, fitting 
perfectly the data. Although the average {\it appears} to be constant for 
low noise, it depends, in fact, on the noise {\it for all $D$}. The same 
applies also to nonlinear systems (cf. fit in  Fig. 1b) and {\it bounded}
noise.   
\\
\indent
(iv) Because the value of the threshold is arbitrary, any scaling 
can be achieved, just by tuning $D_c$, as fittings of the form of Eq. (1) of
\cite{Lai:2002} are very sensitive to the value of $D_c$. To verify this, 
we show in Fig. 1b the logistic map $x_{n+1} = a x_n (1 - x_n) + D\,\xi_n$  
with $\langle  \xi_n,\xi_m \rangle = \delta_{nm}$ as in \cite{Lai:2002}. 
The putative threshold of $D_c = 10^{-5}$ is marked by an arrow. 
The graph is evidently not constant there.
With $D_c \approx 7 \cdot 10 ^{-6}$ (first arrow), a scaling $\Delta G \sim 
(D - D_c)^\alpha$ results, with $\Delta G$ of \cite{Lai:2002}. This
yields $\alpha \approx 2.1$ (Fig. 1c), in clear contrast to $\alpha \approx
1$ reported in \cite{Lai:2002}. Therefore, no reliable (i. e. any) exponent
can be obtained, since $D_c$ is not well defined from the outset. 
\begin{figure}[h]
\begin{center}
\includegraphics[angle=0,width=2.6cm,height=2.6cm]{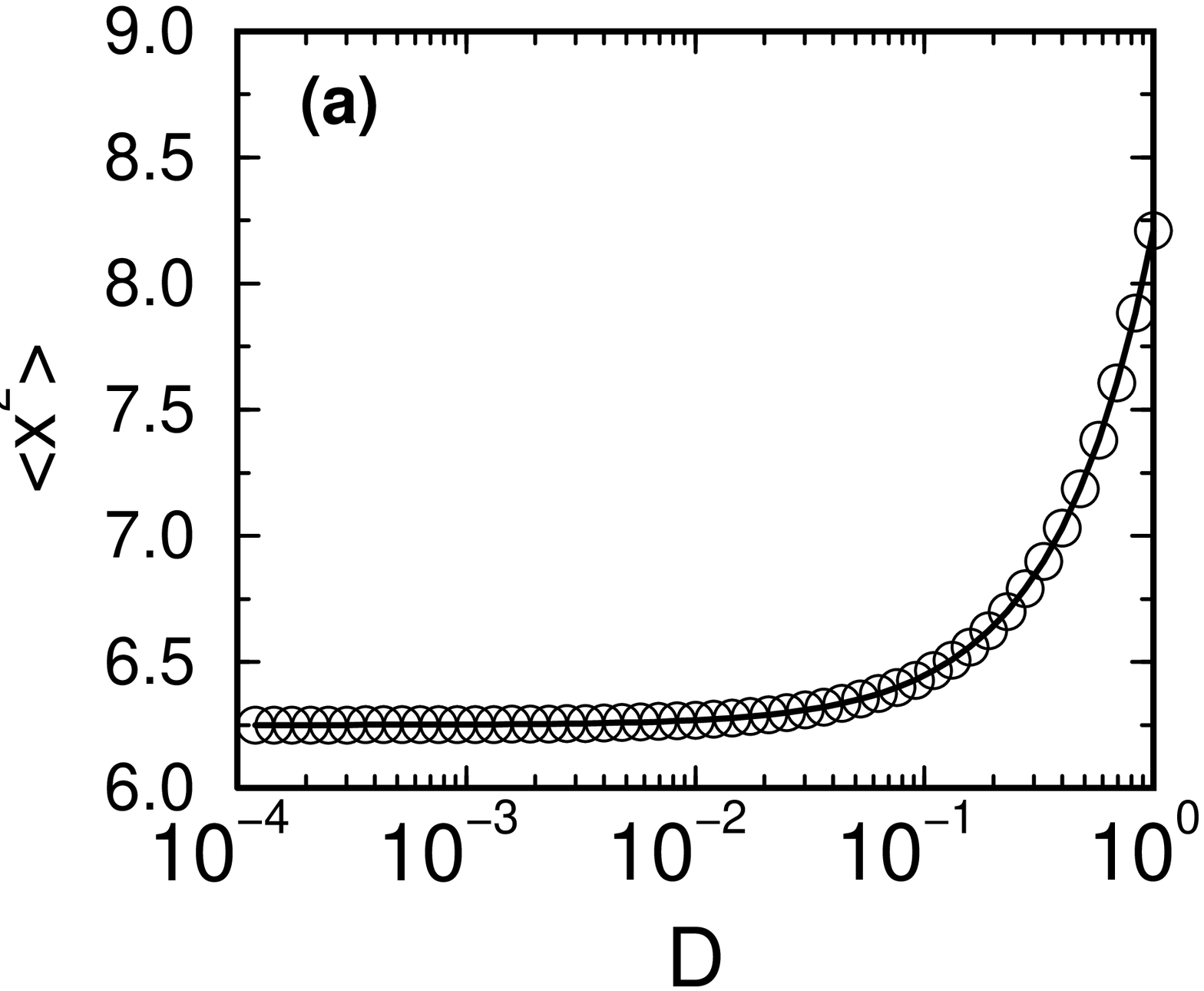}
\includegraphics[angle=0,width=2.6cm,height=2.6cm]{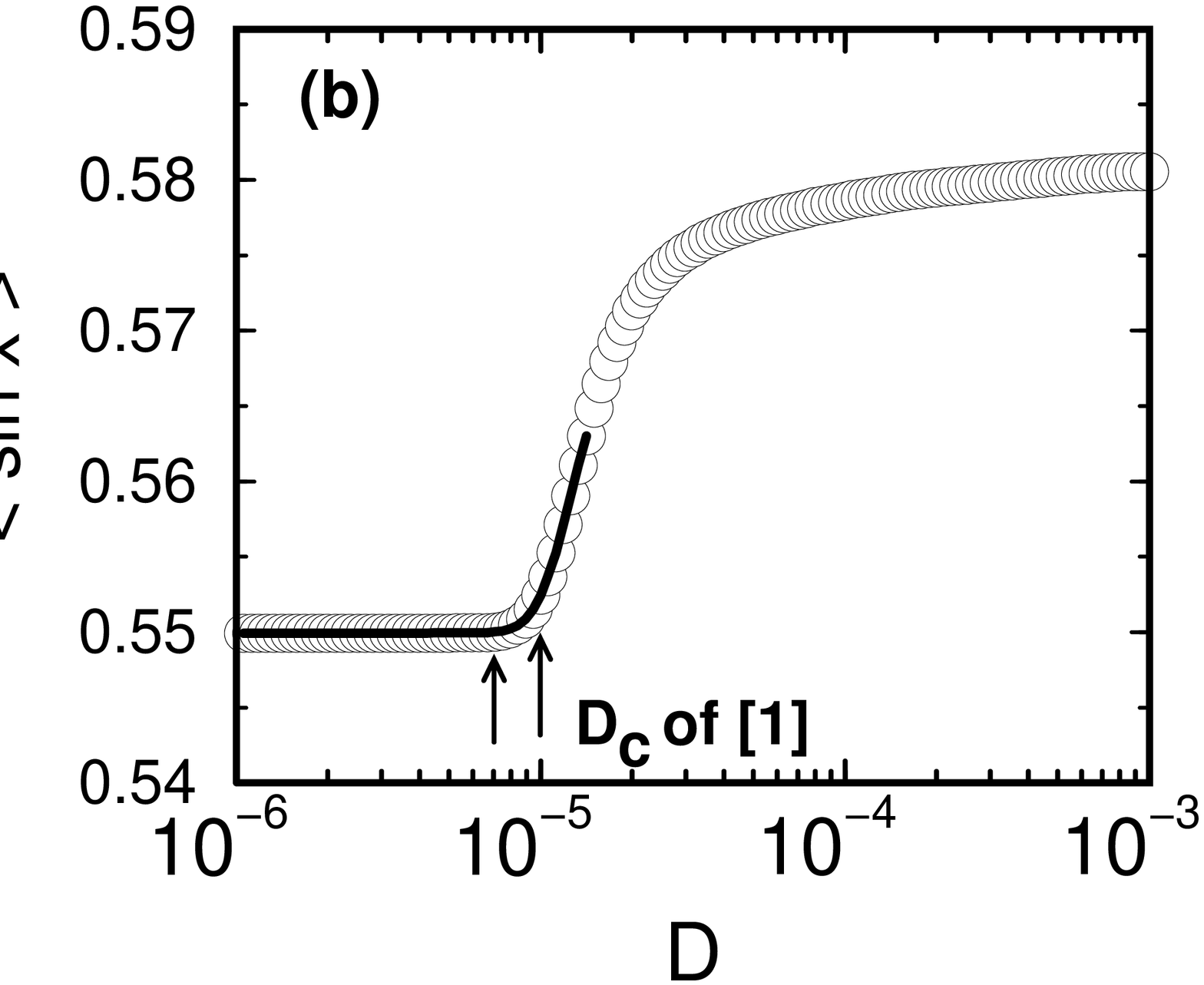}
\includegraphics[angle=0,width=2.6cm,height=2.6cm]{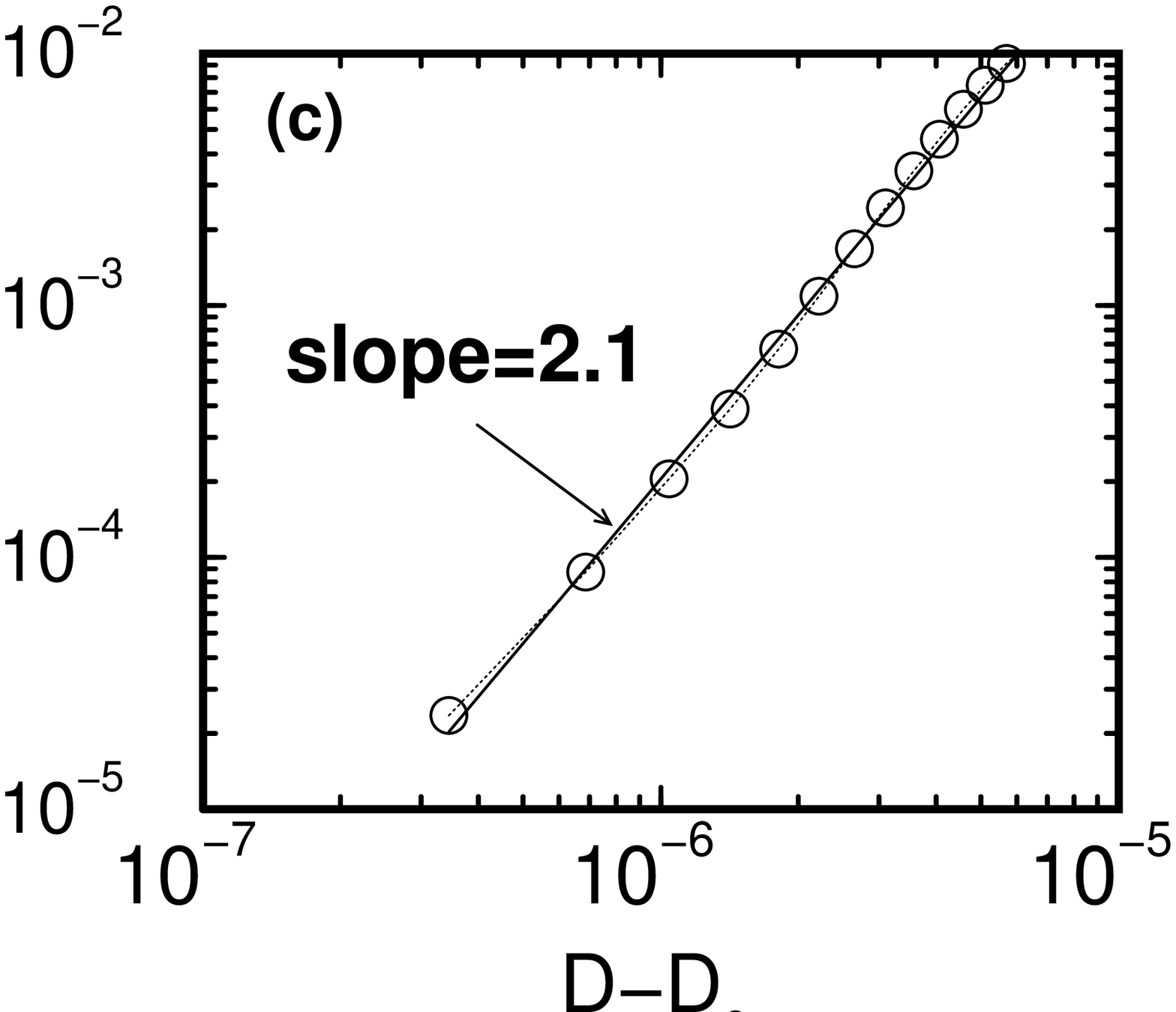}
\caption{(a) $\langle x^2 \rangle$ for the linear map vs. $D$ with 
$a=0.7$ and $b=0.75$ (circles) and analytical fit $\langle x^2 \rangle =
x_{\star}^2 + \frac{1}{1-a^2} \, D$ (full line). (b) $\langle \sin\,x \rangle$
for the logistic map with $a=3.8008$ and a polynomial fit for $D\, \le\,
1.4 \cdot 10 ^{-5}$ (full line) (c) Scaling $\Delta G \sim (D - D_c)^\alpha$
with $D_c \approx 7 \cdot 10 ^{-6}$ (circles) and least squares fit (full
line). All averages are for $10^{8}$ iterations.} 
\label{average}
\end{center}
\end{figure}
\\
\indent
As to the reply, the authors claim to have ``a theoretical justification
for the existence of a threshold'' through $D_c = \sqrt{{\Delta \Phi}/\ln 
\chi^{-1}}$, with ${\Delta\Phi}$ the quasipotential (see (ii)) and $\chi$ 
the probability resolution. Contrary to Ref. [5] of the reply, where $\chi$ 
drops out since only proportionalities under parameter variation are
considered,  $D_c$ in \cite{Lai:2002} depends for fixed parameters on 
$\chi$ and hence $\alpha$ is not uniquely defined. Applied to (iv) with 
$D_c=10^{-5}$ of \cite{Lai:2002} and ${\Delta\Phi} \approx7.8\cdot10^{-10}$ 
(not shown) yields $\chi=0.0004$. But a finer resolution $\chi = 10^{-7}$ 
gives $D_c=7\cdot10^{-6}$, with  $\alpha \approx 2.1$ (cf. (iv)), whereas 
$\chi = 10^{-2}$ results in $D_c=1.3\cdot10^{-5}$ and $\alpha \approx 0.21$ 
(not shown). Again, this invalidates any proper scaling. 
\\
\\
\newline
Suso Kraut\\
\indent {\small Instituto de F{\'i}sica, Universidade de S\~ao Paulo} \\
\indent {\small Caixa Postal 66318, 05315-970 S\~ao Paulo, Brazil}\\
\\
\\

\end{document}